\documentclass[final]{aipproc}

\layoutstyle{6x9}

\usepackage{graphicx}
\newenvironment{palatino}{\fontencoding{OT1}\fontfamily{ppl}\selectfont}{}
\newenvironment{compacttext}{\begin{minipage}[t]{\textwidth}\footnotesize}{\end{minipage}}

\begin{document}

\title{Relations between the Gribov-horizon and center-vortex confinement scenarios\thanks{Plenary talk
presented by \v{S}.\ Olejn{\'\i}k at \emph{Quark Confinement and the
Hadron Spectrum VI\/}, Villasimius, Sardinia, Italy, Sep.\ 21--25,
2004.}}

\author{Jeff Greensite}{
  address={Physics and Astronomy Dept., San Francisco State
University, San Francisco, CA~94117, USA},
  email={greensit@sfsu.edu} }

\author{\v{S}tefan Olejn{\'\i}k}{
  address={Institute of Physics, Slovak Academy
of Sciences, SK--845 11 Bratislava, Slovakia},
  email={stefan.olejnik@savba.sk} }

\author{Daniel Zwanziger}{
  address={Physics Department, New York University, New York, NY~10003,
  USA},
  email={daniel.zwanziger@nyu.edu} }

\begin{abstract}
We review numerical evidence on connections between the
center-vortex and Gribov-horizon confinement scenarios.

\end{abstract}

\maketitle


\begin{flushright}
~\begin{minipage}[t]{0.45\textwidth}
\begin{flushright}
\begin{palatino}{\footnotesize
\emph{So oft in theologic wars,\\
~\hspace*{3mm}The disputants, I ween,\\
Rail on in utter ignorance\\
~\hspace*{3mm}Of what each other mean,\\
And prate about an Elephant\\
~\hspace*{3mm}Not one of them has seen!}\\[0.2mm]
\footnotesize \emph{\textbf{John Godfrey Saxe}} (1816-1887)
\cite{Saxe:1887}}
\end{palatino}\end{flushright}
\end{minipage}
\end{flushright}


\section{Introduction}\label{intro}

    Once upon a time, several blind men from a Hindustan village touched
different parts of an elephant's body and, judging from their
perceptions, quarreled about what the whole creature might look
like. In a similar way, we explore color confinement by isolating
different aspects of the phenomenon, and build our limited
experience into various model schemes. The moral from the ancient
parable teaches us that it is vitally important to unify all
different views before we can appreciate the whole beauty of the
beast.

    The aspirations of this contribution are more modest. I will
concentrate on common points of two seemingly unrelated pictures of
color confinement. The former assumes that confinement arises due to
the condensation of a particular type of topological excitations, so
called center vortices (for a review see \cite{Greensite:2003bk}),
in the QCD vacuum; while the latter is the Gribov-horizon scenario
in Coulomb gauge, which has been advocated by Gribov
\cite{Gribov:1977wm} and Zwanziger \cite{Zwanziger:1998ez}.

    I will first briefly introduce the idea of the Gribov-horizon
scenario, then formulate a simple criterion of confinement for
static color charges in Coulomb gauge, discuss how the fulfillment
of this criterion depends on presence/absence of center vortices,
and finally present results for the Coulomb energy of a pair of
static charges. I will present a subset of our numerical results,
details, as well as some analytic insights on the connections
between center vortices and the Gribov horizon, can be found in
recent publications
\cite{Greensite:2003xf,Greensite:2004ke,Greensite:2004ur}.


\section{Confinement scenario in Coulomb gauge}\label{gribov}

    In Coulomb gauge, the Hamiltonian of QCD has the following form \cite{Christ:1980ku}:
\begin{eqnarray}\label{hamiltonian}
H&=&H_{glue}+H_{coul},\\
H_{glue}&=&\frac{1}{2}\int d^3\mathbf{x}
\left({\cal{J}}^{-1/2}\mathbf{E}^a{\cal{J}}\cdot\mathbf{E}^a
{\cal{J}}^{-1/2}+\mathbf{B}^a\cdot\mathbf{B}^a\right),\\
H_{coul}&=&\frac{1}{2}\int d^3\mathbf{x}d^3\mathbf{y}\;
{\cal{J}}^{-1/2}\varrho^a(x){\cal{J}}K^{ab}(x,y;A)\varrho^b(y){\cal{J}}^{-1/2},\\
\varrho^a&=&\varrho^a_{matter} -g
f^{abc}\mathbf{A}^b\cdot\mathbf{E}^c.
\end{eqnarray}
A prominent role in this expression is played by the Faddeev--Popov
operator
\begin{equation}\label{FPOperator}
M(A)\equiv-\nabla\cdot{\cal{D}}(A), \qquad
\mbox{where}\qquad{\cal{D}}^{ac}_i(A)=\partial_i\delta^{ac}+f^{abc}A^b_i(x),
\end{equation}
which enters both the interaction kernel $K$ and the Jacobian factor
$\cal{J}$
\begin{equation}\label{KandJ}
K^{ab}(x,y;A) \equiv \left[M^{-1}
\;(-\nabla^2)\;M^{-1}\right]^{a,b}_{x,y}, \qquad
{\cal{J}}\equiv\det\left[M(A)\right].
\end{equation}

    It is well-known since the seminal paper of Gribov \cite{Gribov:1977wm}, that
Coulomb-gauge fixing in a non-abelian theory is a difficult problem.
The transversality condition $\nabla\cdot\mathbf{A}=0$ does not fix
the gauge completely. Gribov \cite{Gribov:1977wm} suggested to
restrict integration over configurations to the so-called Gribov
region (GR), defined as the subspace of transverse gauge fields for
which the Faddeev--Popov operator is positive, and which are
therefore local minima of the functional
\begin{equation}\label{functional}
I[\mathbf{A},g]=\int dx\left[^g \mathbf{A}^a(x)\right]^2,\qquad
\mbox{where}\qquad{}^gA_i=g^{-1} A_i g +g^{-1}\partial_i g.
\end{equation}
The boundary of this region is called the Gribov horizon.

    However, there exist gauge orbits which intersect the Gribov
region more than once; the next step \cite{Dell'Antonio:1991xt} then
is to restrict fields to the fundamental modular region (FMR), the
set of absolute minima of the functional (\ref{functional}). Both
the GR and the FMR are bounded in every direction and convex.

    The essence of the Gribov-horizon confinement scenario can be
phrased in a simple way: The dimension of gauge-field configuration
space is huge, so it is reasonable to expect that most
configurations are located close to its boundary (the horizon; in a
similar way, the volume measure $r^{d-1}dr$ of a $d$-dimensional
sphere is peaked at its surface). The interaction kernel $K$, which
determines the interaction energy of static color sources, contains
the inverse of the Faddeev--Popov operator, which is strictly zero
on the horizon and near-zero close to the horizon. A high density of
configurations near the horizon can thus lead to a strong
enhancement of the Coulomb interaction energy, and hopefully cause
color confinement.


\section{A confinement condition in terms of Faddeev--Popov eigenstates}\label{criterion}

    Let us consider a single static color charge, which can be written in
Coulomb gauge as
\begin{equation}\label{staticcharge}
\Psi_C^\alpha[A;x]=\psi^\alpha(x)\Psi_0[A],
\end{equation}
where $\alpha$ is the color index for a point charge in color group
representation $r$, and $\Psi_0$ is the Coulomb-gauge ground state.
The excitation energy of this state, above the ground state, is
given by
\begin{equation}\label{E_r}
{\cal{E}}_r =\frac{\langle \Psi^\alpha_C \vert H_{coul} \vert
\Psi^\alpha_C \rangle} {\langle \Psi^\alpha_C \vert \Psi^\alpha_C
\rangle} - \langle \Psi_0 \vert H_{coul} \vert \Psi_0 \rangle,
\end{equation}
and is proportional to
\begin{equation}\label{E}
{\cal{E}}=\frac{1}{N^2 -1} \langle K^{aa}(x,x;A) \rangle.
\end{equation}
The quantity ${\cal{E}}_r$ is the color Coulomb self-energy of
unscreened color charge and is expected to be both ultraviolet and
infrared divergent in a confining theory. The UV divergence can be
regulated by a lattice cut-off, however, the quantity must still be
divergent at infinite volume, even after lattice regularization, due
to IR effects.

    On a lattice, one can express (\ref{E}) through eigenstates of the Faddeev--Popov operator
\begin{equation}\label{eigenequation}
{\sum_{b,y}}M^{ab}_{xy}\;\phi^{(n)b}_{y}=\lambda_n\phi^{(n)a}_{x}
\end{equation}
simply as (assuming that $M$ is invertible, i.e.\ excluding its zero
modes)
\begin{equation}\label{Ethrueigenstates}
{{\cal{E}}=\frac{1}{3V_3}\sum_n
\left\langle\frac{F_n}{\lambda_n^2}\right\rangle},
\qquad\mbox{where}\qquad F_{n}\equiv{\displaystyle{\sum_{a,xy}
\phi^{(n)a}_{x} (-\nabla^2)_{xy}\phi^{(n)a\ast}_{y}}},
\end{equation}
and $V_3$ is the lattice 3-volume. In SU(2) lattice gauge theory,
the link variables can be expressed as
\begin{equation}\label{U}
U_\mu(x)=b_\mu(x)+i\sigma^c a_\mu^c(x), \qquad b_\mu(x)^2+\sum_c
a_\mu^c(x)^2=1,
\end{equation}
and the lattice Faddeev--Popov operator
\begin{eqnarray}\label{latticeFP}
\nonumber
  M^{ab}_{xy} &=& \delta^{ab} \sum_{k}\left\{ \delta_{xy}
\left[b_k(x)
   + b_k(x-\hat{k})\right] - \delta_{x,y-\hat{k}} b_k(x)
   -  \delta_{y,x-\hat{k}} b_k(y) \right\}\\
   &-& \epsilon^{abc} \sum_{k} \left\{ \delta_{x,y-\hat{k}}
a^c_k(x) - \delta_{y,x-\hat{k}} a^c_k(y)  \right\}
\end{eqnarray}
($x,y$ denote lattice sites at fixed time) is a $3V_3\times 3V_3$
sparse matrix with $3V_3$ linearly independent eigenstates. If we
denote $N(\lambda,\lambda+\Delta\lambda)$ the number of eigenvalues
in the range between $\lambda$ and $\lambda+\Delta\lambda$, we can
introduce, on a large lattice, the density of states
$\rho(\lambda)\equiv
N(\lambda,\lambda+\Delta\lambda)/(3V_3\Delta\lambda)$. Then, as
$V_3\to\infty$,
\begin{equation}\label{Eintegral}
{\cal{E}}=\int_0^{\lambda_{max}}\frac{d\lambda}{\lambda^2}\left\langle\rho(\lambda)F(\lambda)\right\rangle.
\end{equation}
A (necessary) condition for confinement can now be formulated:
\emph{The excitation energy ${\cal{E}}_r$ of a static, unscreened
color charge is divergent if, at infinite volume,}
\begin{equation}\label{condition}
\lim_{\lambda\to0}\frac{\left\langle\rho(\lambda)F(\lambda)\right\rangle}{\lambda}>0.
\end{equation}
%


\section{Center vortices and the confinement condition}\label{vortices}

    It is interesting to check whether the above condition is fulfilled in
various ensembles of lattice configurations. First, at zero-th order
in the gauge coupling, the Faddeev--Popov operator is simply a
lattice Laplacian and its eigenstates are just plane waves. One can
easily verify that, in this case,
\begin{equation}\label{free}
\rho(\lambda)=\frac{1}{4\pi^2}\sqrt{\lambda}, \qquad
F(\lambda)=\lambda, \qquad
{\cal{E}}=\frac{1}{2\pi^2}\sqrt{\lambda_{max}},
\end{equation}
the excitation energy is IR finite and the confinement criterion is
not met. One needs some mechanism of enhancement of $\rho(\lambda)$
and $F(\lambda)$ in the region of small $\lambda$ values. I~will
argue that such an enhancement exists in full lattice
configurations, and is provided by center vortices.

    Center vortices are identified by fixing to an adjoint gauge,
and then projecting link variables to the $Z_N$ subgroup of SU($N$)
\cite{DelDebbio:1996mh}. The excitations of the projected theory are
known as P-vortices. In the direct maximal center gauge (DMCG) in
SU(2) \cite{DelDebbio:1998uu} one fixes to the maximum of
\begin{equation}\label{DMCG}
{\cal{R}}_{MCG}=\sum_{x,\mu}
\left\vert\textstyle{\frac{1}{2}}\mbox{Tr}[U_\mu(x)] \right\vert^2,
\end{equation}
and center projects by
\begin{equation}\label{centerprojection}
U_\mu(x) \longrightarrow Z_\mu(x) = \mbox{sign Tr}[U_\mu(x)].
\end{equation}
A lot of evidence has been accumulated that center vortices alone
reproduce much of confinement physics, for a review see
\cite{Greensite:2003bk}.

    We have determined the density of states $\rho(\lambda)$ and the
mean value of the lattice Laplacian $F(\lambda)$ in three ensembles
of lattice configurations:
\begin{enumerate}
    \item full configurations, $\lbrace U_\mu(x)\rbrace$,
    \item ``vortex-only'' configurations, $\lbrace Z_\mu(x) = \mbox{sign
    Tr}[U_\mu(x)]\rbrace$, and \phantom{$Z_\mu^\dagger(x)$}
    \item ``vortex-removed'' configurations, $\lbrace U_\mu^{(R)}(x)
= Z_\mu^\dagger(x) U_\mu(x)\rbrace$.
\end{enumerate}
\begin{figure}[t!]
\begin{tabular}{c p{0.04\textwidth} c}
  \includegraphics[width=0.47\textwidth]{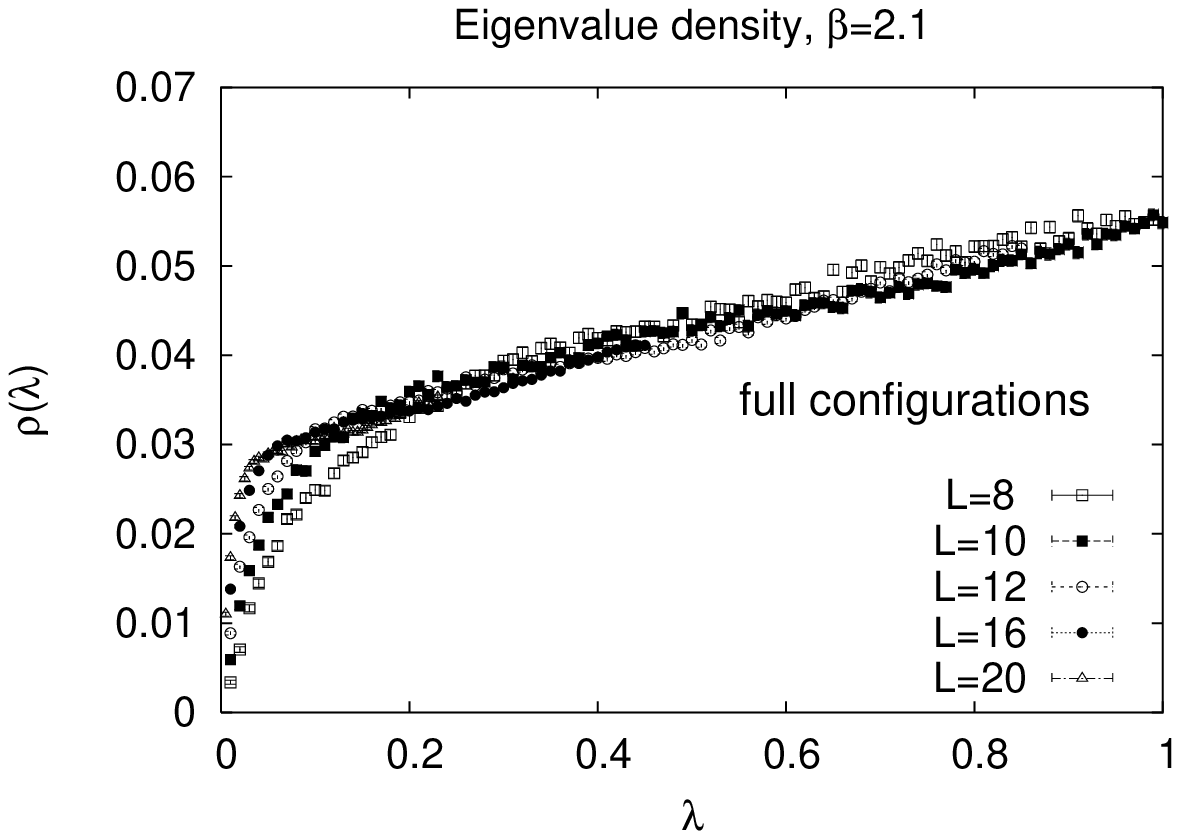}& &
  \includegraphics[width=0.47\textwidth]{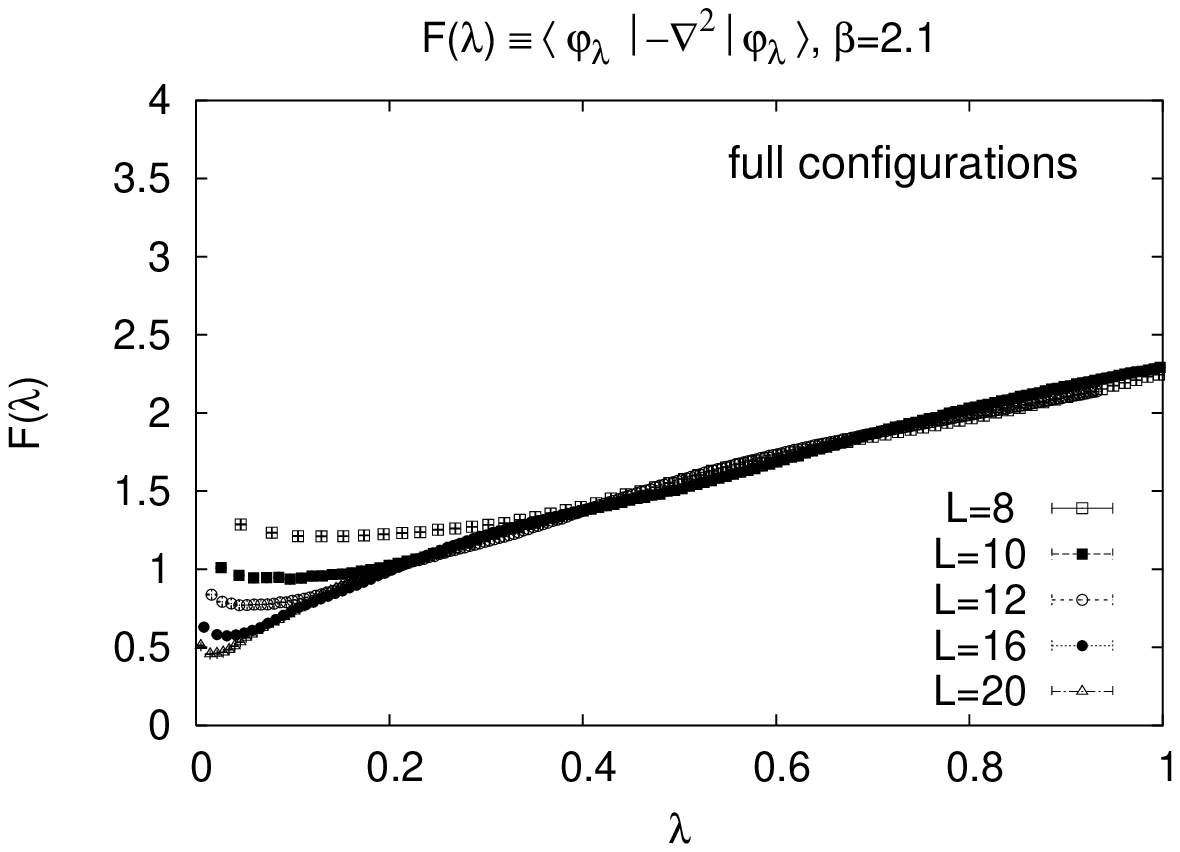}
\end{tabular}
\caption{The density $\rho(\lambda)$ and $F(\lambda)$ for full
lattice configurations.}\label{fig1}
\end{figure}

    The procedure of vortex removal, first introduced by de Forcrand and
D'Elia \cite{deForcrand:1999ms}, is known to remove the string
tension, eliminate chiral symmetry breaking, and send the
topological charge of lattice configurations to zero.

    Each of the three ensembles was brought to Coulomb gauge by
maximizing, on each time-slice,
\begin{equation}\label{coulomb}
{\cal{R}}_{coul}(t)={\sum_{\mathbf{x}}\sum_{k=1}^3}
{\textstyle{\frac{1}{2}}}\mbox{Tr}[U_k(\mathbf{x},t)].
\end{equation}

    Our results for full configurations at $\beta=2.1$ are exemplified in Figure
\ref{fig1}, for a variety of lattice volumes. One can observe a
sharp ``bend'' near $\lambda\to0$. Both quantities behave near 0
like $\lambda^p$, $\lambda^q$, with $p,q$ small numbers. A scaling
analysis along the lines of random matrix theory (see Appendix B in
\cite{Greensite:2004ur} for details and references) gives the
estimates
\begin{equation}\label{powersfull}
\rho(\lambda)\sim\lambda^{0.25},\qquad F(\lambda)\sim\lambda^{0.38},
\end{equation}
with a subjective error of about 20\% in the exponents. The Coulomb
gauge confinement condition, Eq.\ (\ref{condition}), is thus clearly
fulfilled. This result provides a confirmation of the mechanism
envisaged in the Gribov-horizon scenario.

\begin{figure}[b!]
\begin{tabular}{c p{0.04\textwidth} c}
  \includegraphics[width=0.47\textwidth]{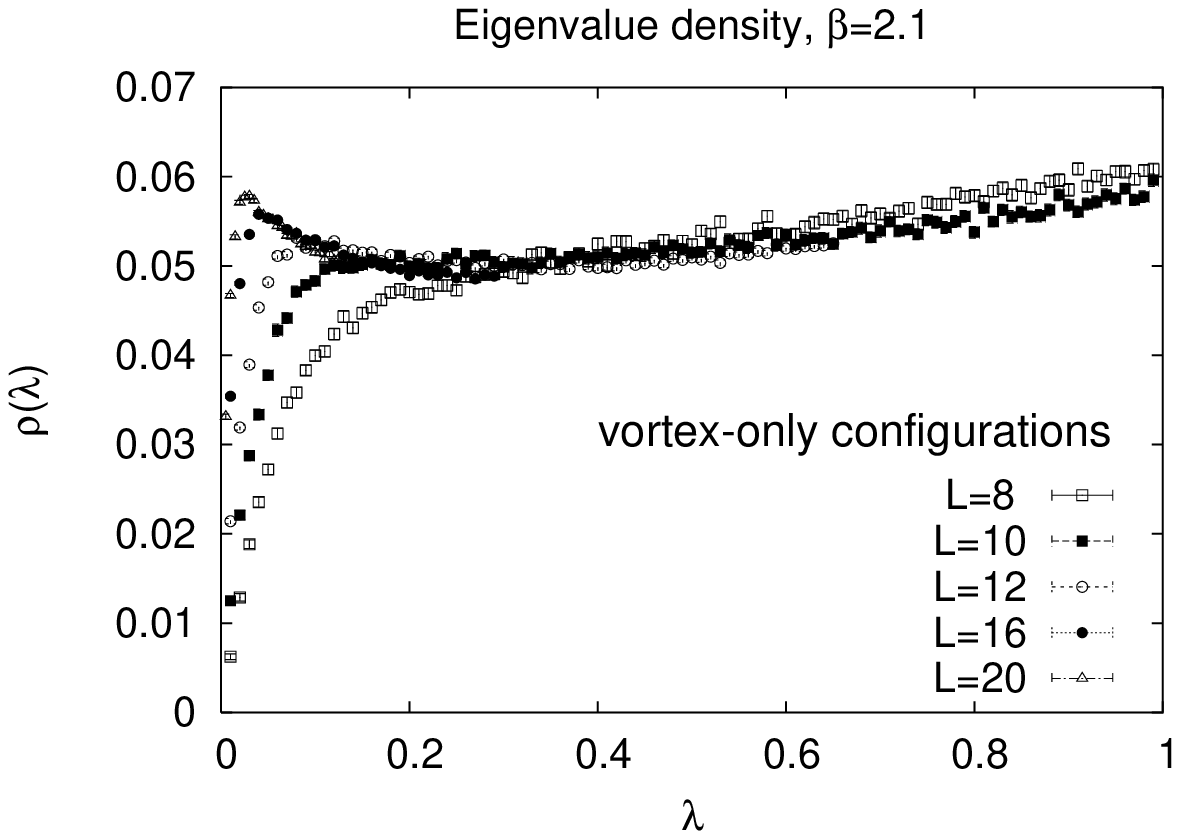}& &
  \includegraphics[width=0.47\textwidth]{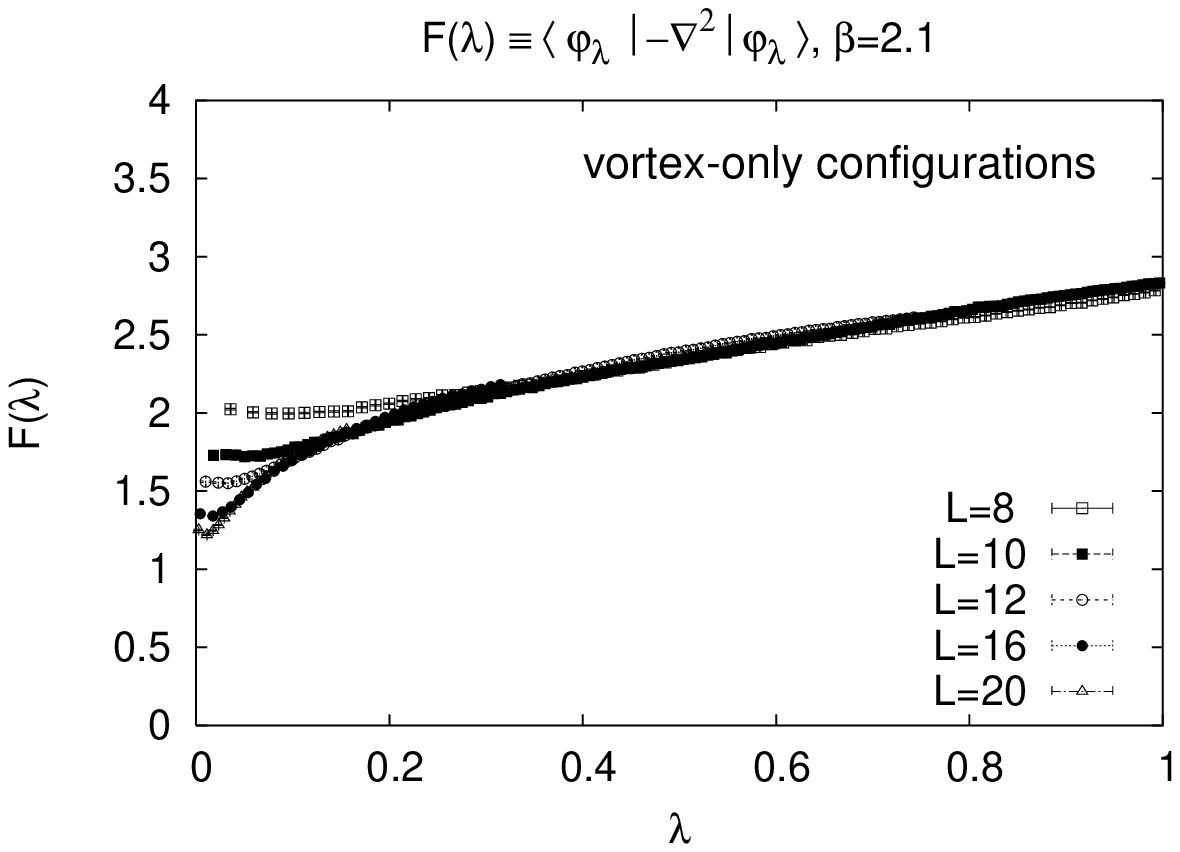}
\end{tabular}
\caption{The density $\rho(\lambda)$ and $F(\lambda)$ for
vortex-only configurations.}\label{fig2}
\end{figure}

    To pinpoint the mechanism which might be responsible for the
enhancement of eigenvalues near the horizon, we considered the
Faddeev--Popov observables in the vortex-only configurations. Our
data are displayed in Figure \ref{fig2}. The enhancement near 0 is
even more pronounced; it appears that both quantities converge to a
non-zero value in the infinite volume limit
\begin{equation}\label{powerscp}
\rho(0)\sim 0.06,\qquad F(0)\sim 1.0,
\end{equation}
which is confirmed also by a RM scaling analysis. Once again, the
confinement criterion (\ref{condition}) is obviously satisfied.

\begin{figure}[t!]
\begin{tabular}{c p{0.04\textwidth} c}
  \includegraphics[width=0.47\textwidth]{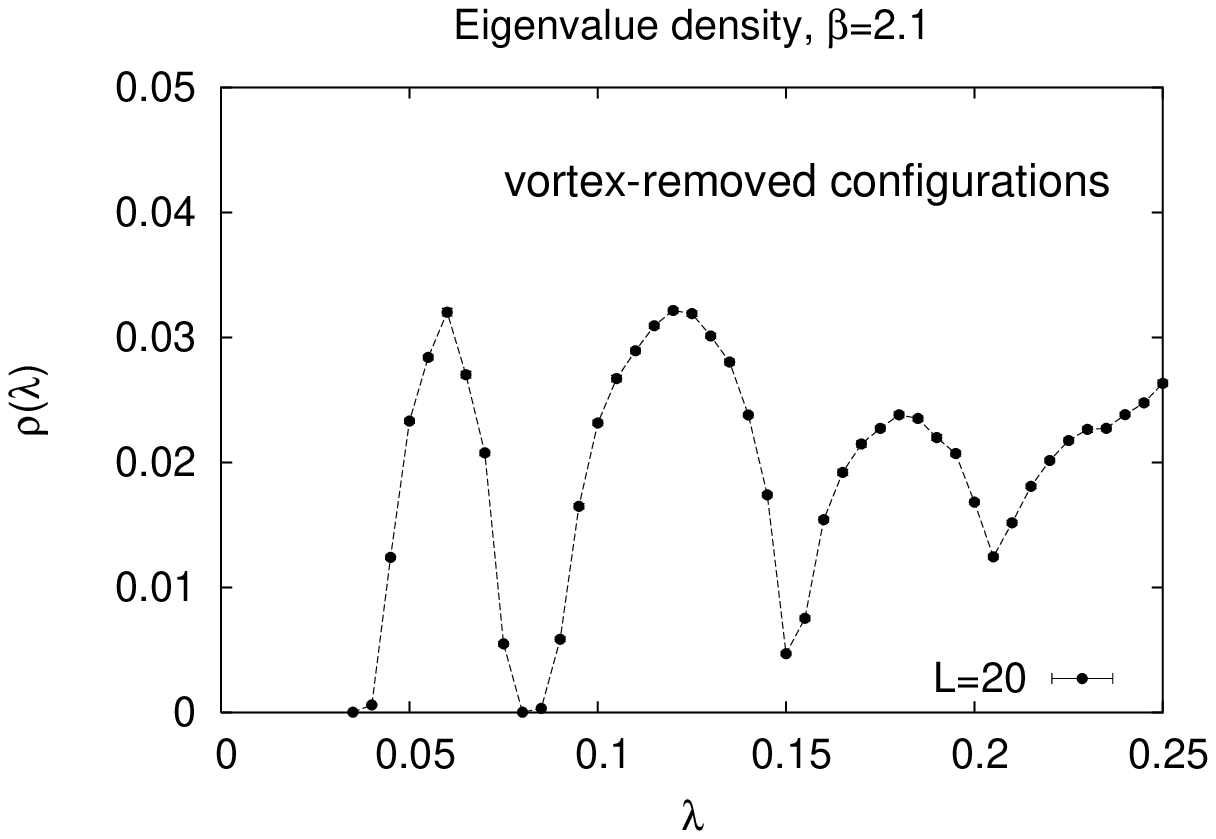}& &
  \includegraphics[width=0.47\textwidth]{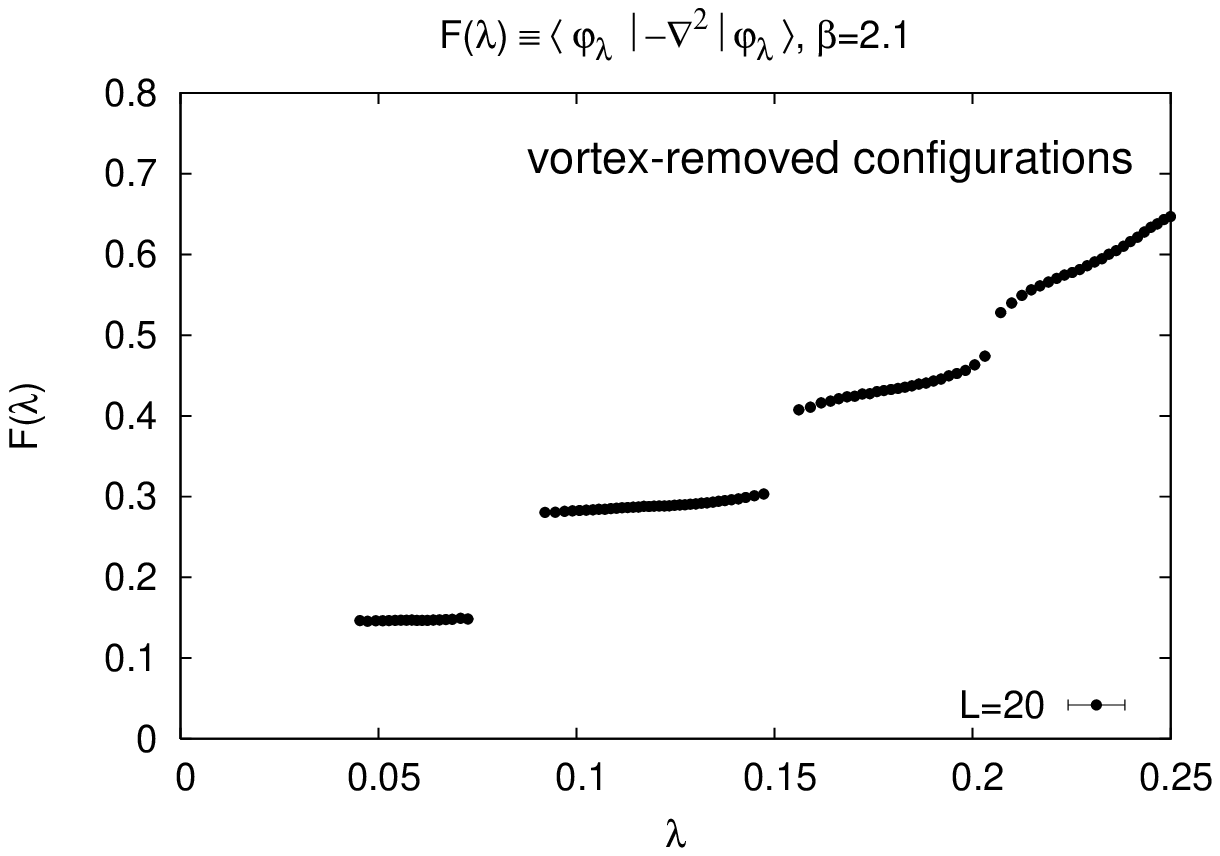}
\end{tabular}
\caption{The density $\rho(\lambda)$ and $F(\lambda)$ for
vortex-removed configurations.}\label{fig3}
\end{figure}
\begin{figure}[b!]
\begin{tabular}{c p{0.04\textwidth} c}
  \includegraphics[width=0.47\textwidth]{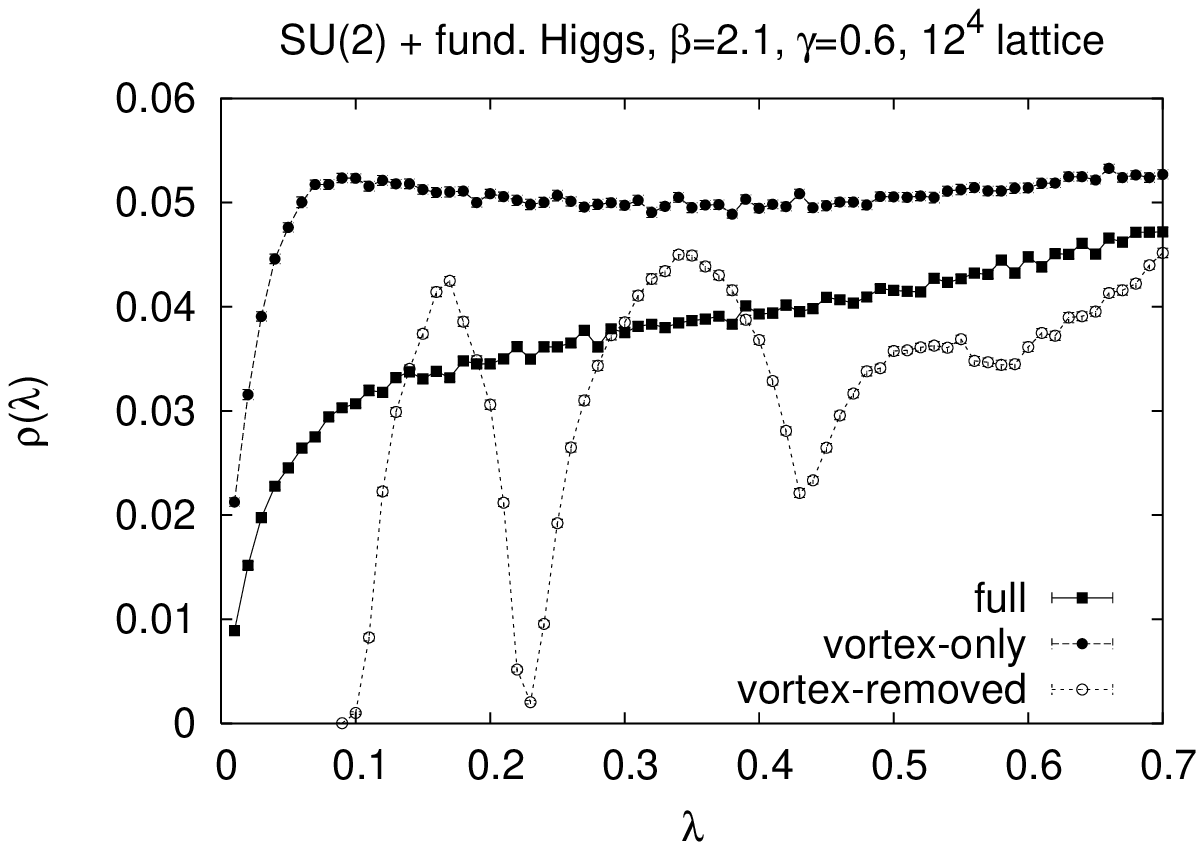}& &
  \includegraphics[width=0.47\textwidth]{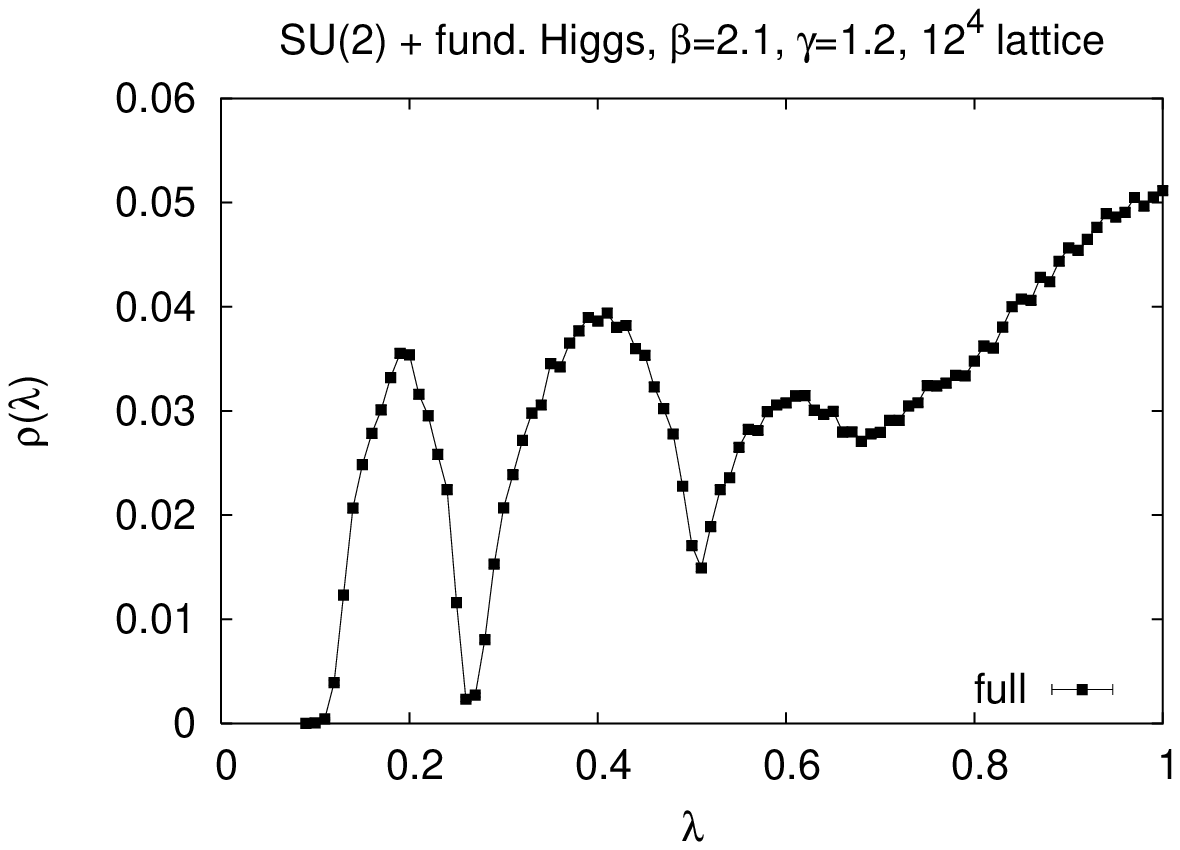}
\end{tabular}
\caption{The density $\rho(\lambda)$ for a gauge-Higgs system in the
``confined'' phase (left) and the ``Higgs'' phase
(right).}\label{fig4}
\end{figure}

    Finally, results for ``vortex-removed'' configurations are shown
in Figure \ref{fig3}, for the largest available, $20^4$ lattice
volume. The eigenvalue spectrum is strikingly different in this
case. The eigenvalue density consists of a series of distinct peaks,
while values of $F(\lambda)$ are organized into separate bands, the
lowest few of them clearly separated by gaps. This result can easily
be understood by inspection of the eigenvalue spectrum of the
Laplacian operator (equal to $M$ at zero-th order in the coupling).
The eigenvalue density, at finite volume, is a sum of
delta-functions, and each eigenvalue $\lambda_k$ is multiply
($N_k$-times) degenerate. The quantity $F(\lambda_k)$ is equal to
$\lambda_k$. Now if one compares the degeneracy $N_k$ of the $k$-th
eigenvalue with the number of eigenvalues inside the $k$-th ``band''
of $F(\lambda)$ (the right panel of Figure \ref{fig3}), one finds a
precise match. This can be simply interpreted: the vortex-removed
configuration is just a small perturbation of the zero-field limit
$U_\mu=\mathbf{1}$. This perturbation lifts the degeneracy of
degenerate eigenvalues and spreads them into bands of finite width.
The estimate of the Coulomb self-energy in this case indicates that
it remains IR finite in the infinite volume limit.

    In pure gauge theory, we used the procedure of de Forcrand and
D'Elia \cite{deForcrand:1999ms} to study effects of vortex removal
on the density of Faddeev--Popov eigenvalues in the small $\lambda$
region. However, in some lattice models, confining vortex
configurations can be suppressed by changing the coupling constants.
A prototype example is the gauge field coupled to a scalar field of
unit modulus in the fundamental representation of the gauge group
\cite{Lang:1981qg}. The action of the model is
\begin{equation}\label{su2higgs}
S = S_W + {\textstyle\frac{\gamma}{2}} \sum_{x,\mu}
\mbox{Tr}[\phi^\dagger(x)U_\mu(x)\phi(x+\widehat{\mu})],
\end{equation}
where $S_W$ is the usual Wilson gauge action and $\phi$ is an SU(2)
group-valued field. In the strict sense, this theory is
non-confining for all values of couplings $\beta, \gamma$, due to
the well-known Osterwalder-Seiler--Fradkin-Shenker theorem
\cite{Osterwalder:1977pc,Fradkin:1978dv}. Even though there is no
thermodynamic phase transition separating the pseudo-Higgs phase
from the pseudo-confinement phase of the model, there is a symmetry
breaking transition between the phases (a ``Kert\'esz'' line), and
center vortices percolate in the ``confinement'' phase, and cease to
percolate in the ``Higgs'' phase
\cite{Greensite:2004ke,Chernodub:1998wh,Langfeld:2001he,Bertle:2003pj}.
By varying $\gamma$ at fixed gauge coupling $\beta$, one can modify
the vortex content and study its effects on Faddeev--Popov
eigenvalues.

    In Figure \ref{fig4} I show the eigenvalue density at
$\beta=2.1$ for two values of gauge-Higgs coupling: $\gamma=0.6$,
deep in the ``confinement'' phase, and $\gamma=1.2$ corresponding to
the ``Higgs'' phase. In the former, the densities for full,
vortex-only, and vortex-removed configurations clearly resemble
those in pure gauge theory, while the density in the Higgs phase for
full configurations looks almost identical to the vortex-removed
data in the confinement phase.


\section{Coulomb energy}\label{coulombenergy}

    We have seen that the Coulomb self-energy of a color
non-singlet state is IR divergent, due to the enhanced density of
Faddeev--Popov eigenvalues near zero. Another question is whether
the color Coulomb potential of a charge-anticharge pair grows
linearly and, if so, whether it is also sensitive to the
presence/absence of center vortices. This question was addressed in
Ref.\ \cite{Greensite:2003xf}, and I will summarize here briefly the
main results.

\begin{figure}[t!]
\begin{tabular}{c}
  \includegraphics[width=0.5\textwidth]{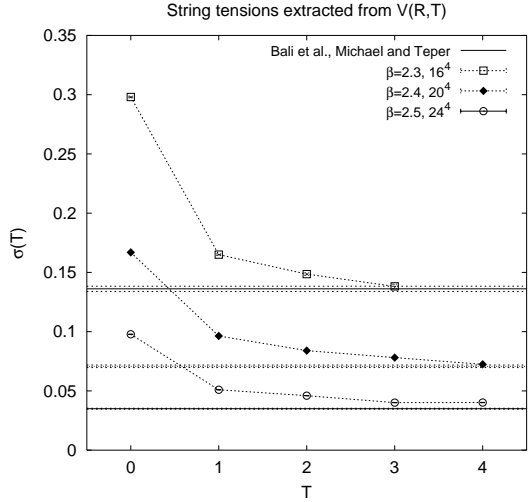}
\end{tabular}
\caption{$\sigma(T)$ vs.\ $T$. Solid lines indicate the accepted
values of the asymptotic string tension at each $\beta$ value, with
dashed lines indicating the error bars.}\label{fig5}
\end{figure}

    Let
$\vert\Psi_{q\bar{q}}\rangle=\bar{q}^a(0)q^a(R)\vert\Psi_0\rangle$
denote a physical heavy static quark-antiquark state in Coulomb
gauge. Then
\begin{equation}\label{Eqq}
{\cal{E}}_{q\bar{q}}= \langle\Psi_{q\bar{q}}\vert H
\vert\Psi_{q\bar{q}}\rangle -\langle\Psi_0\vert H
\vert\Psi_0\rangle= E_{se}+V_{coul}(R)
\end{equation}
is a sum of self-energy contributions and the $R$-dependent color
Coulomb potential. It can be computed from the correlator of two
timelike Wilson lines in Coulomb gauge:
\begin{equation}\label{G}
G(R,T)=\left.\langle{\textstyle{\frac{1}{2}}}\mbox{Tr}
[L^\dagger(\mathbf{x},T)L(\mathbf{y},T)]\rangle\right\vert_{R=\vert\mathbf{x}-\mathbf{y}\vert};
\qquad
L(\mathbf{x},T)=\exp\left[i\int_0^Tdt\;A_0(\mathbf{x},t)\right].
\end{equation}
It is easy to show that
\begin{equation}\label{energies}
  {\cal{E}} = E_{se}+V_{coul}(R)=\lim_{T\to 0}V(R,T);  \qquad
  {\cal{E}}_{min} = E'_{se}+V(R) = \lim_{T\to\infty}V(R,T),
\end{equation}
where
\begin{equation}\label{VfromG}
V(R,T) =-\textstyle{\frac{\partial}{\partial T}} \log[G(R,T)].
\end{equation}
Above, ${\cal{E}}_{min}$ is the minimal energy of a state containing
two static charges, and $V(R)$ is the static interquark potential.
Since ${\cal{E}}>{\cal{E}}_{min}$, it is clear that if $V(R)$ is
confining, then so is $V_{coul}(R)$, and $V(R)$ is bounded from
above by $V_{coul}(R)$, as first proven by
Zwanziger~\cite{Zwanziger:2002sh}.

    On a lattice we introduce
\begin{equation}\label{Llat}
L(\mathbf{x},T)=U_0(\mathbf{x},a)U_0(\mathbf{x},2a)\dots
U_0(\mathbf{x},T),
\end{equation}
and
\begin{equation}\label{VfromGlat}
V(R,T)=-{\textstyle\frac{1}{a}}\log\left[\frac{G(R,T+a)}{G(R,T)}\right],\quad
V(R,0)=-{\textstyle\frac{1}{a}}\log[G(R,a)],
\end{equation}
from which we obtain an estimate of $V_{coul}(R)$ (exact in the
continuum limit).

\begin{figure}[t!]
\begin{tabular}{c p{0.04\textwidth} c}
  \includegraphics[width=0.47\textwidth]{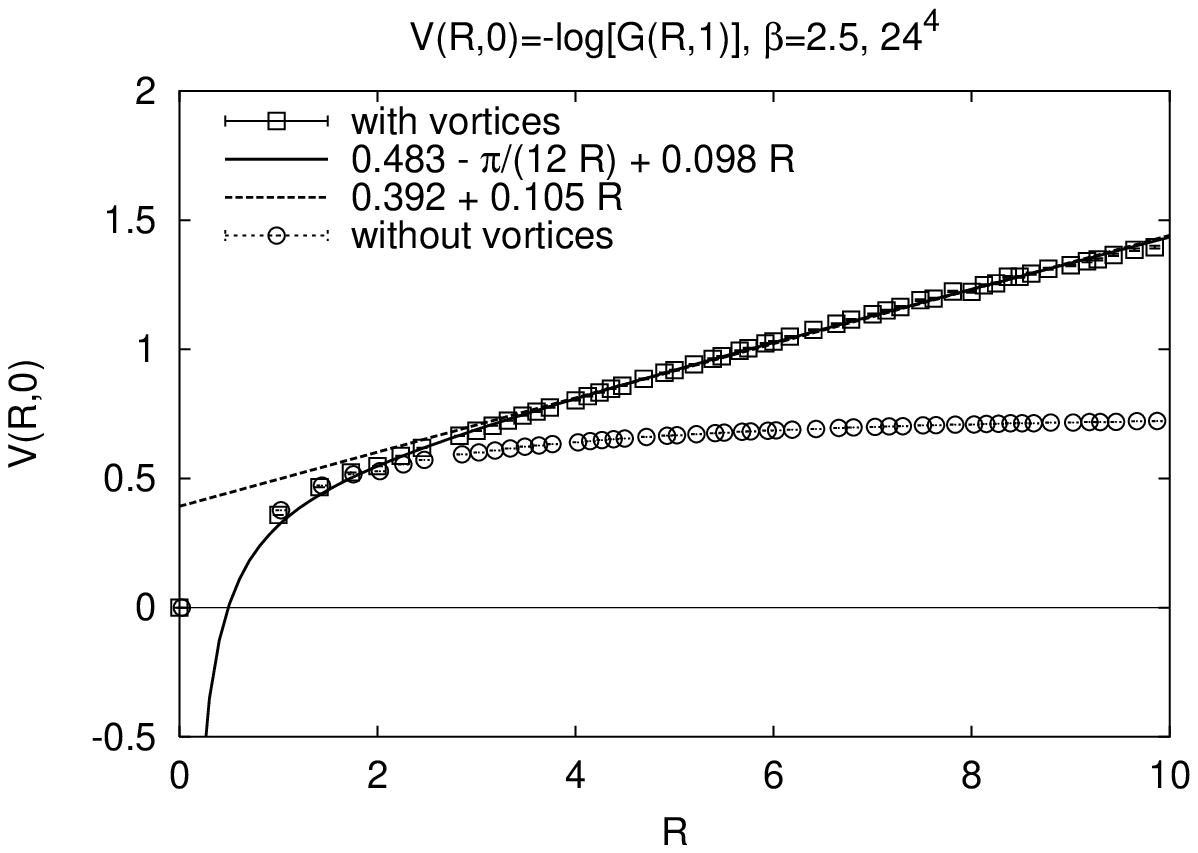}& &
  \includegraphics[width=0.47\textwidth]{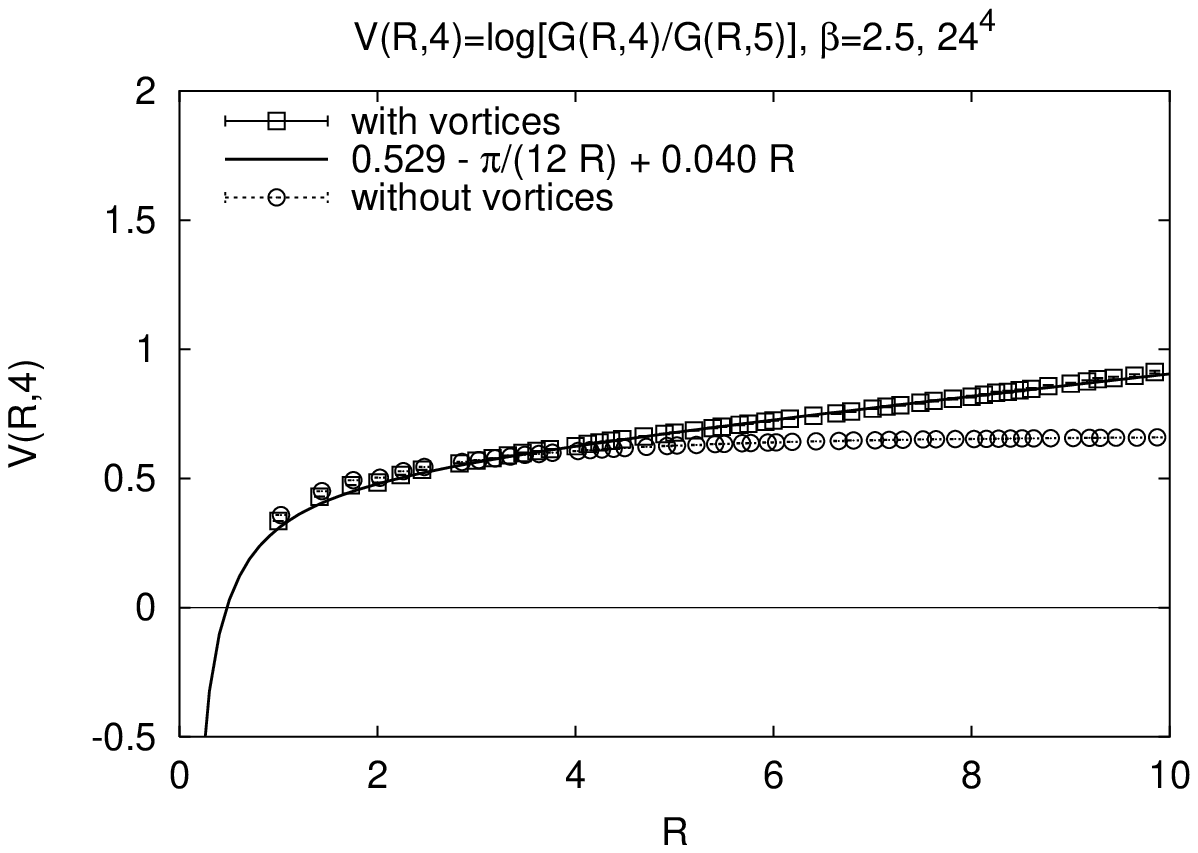}
\end{tabular}
\caption{$V(R,0)$ and $V(R,4)$ at $\beta=2.5$. The former tends to
the color Coulomb potential in the continuum limit, while the latter
approximates the static asymptotic interquark
potential.}\label{fig6}
\end{figure}

    Our results, for SU(2), are shown in Figures \ref{fig5} and
\ref{fig6}.\footnote{First preliminary results of the determination
of the color Coulomb potential in SU(3) lattice gauge theory were
presented by Nakamura \cite{Nakamura:2004an} at this conference.}
The former figure represents a consistency check of our procedure.
It verifies that the string tension $\sigma(T)$ extracted form
$V(R,T)$ approaches the accepted value of the asymptotic string
tension at large $T$. In Figure \ref{fig6} we plot the potential
$V(R,T)$ for $T=0$ (our estimate of the color Coulomb potential) and
for $T=4$ (approaching the static quark-antiquark potential $V(R)$).
Upper lines of data points in both figures clearly demonstrate that
both $V_{coul}(R)$ and $V(R)$ rise linearly with distance. However,
the slope of this linear rise is larger in the color Coulomb
potential, $\sigma_{coul}\approx 3\sigma$.\footnote{The color
Coulomb potential was measured using different methods by Cucchieri
and Zwanziger \cite{Cucchieri:2002su}, and Langfeld and Moyaerts
\cite{Langfeld:2004qs}, with lower values for the ratio
$\sigma_{coul}/\sigma$. The origin of this discrepancy has not been
clarified yet.} The fact that the color Coulomb potential
overconfines does not contradict Zwanziger's bound
\cite{Zwanziger:2002sh}, nor is really surprising. There is no
reason to believe that the quark-antiquark state in the Coulomb
gauge is the true QCD flux tube state, the minimal energy state
containing a static quark and antiquark. The Coulombic force can be
lowered to the true asymptotic one e.g.\ by constituent gluons
present in the QCD flux tube, as was suggested in the gluon chain
model of Greensite and Thorn \cite{Greensite:2001nx}.

    The effect of vortex removal is illustrated in lower lines of
data points of Figure \ref{fig6}. Removing center vortices also
removes the confining property of the color Coulomb potential. Since
the potential, in Coulomb gauge, is sensitive to the interaction
kernel $K$ in the Hamiltonian (\ref{hamiltonian}), and the kernel in
turn to the density of near-zero eigenvalues of the Faddeev--Popov
operator, this result again confirms the observation of the previous
section: Removal of center vortices alters the density quite
drastically.

    We also note that the confining property of the color Coulomb
potential is tied to the unbroken realization of a remnant global
gauge symmetry in Coulomb gauge. This connection was studied in
detail in Ref.\ \cite{Greensite:2004ke}. It was demonstrated there
on a few examples (deconfined phase in pure gauge theory,
pseudo-confinement phase of the gauge--fundamental-Higgs theory)
that confinement in the color Coulomb potential is not identical to
confinement in the static interquark potential, and center symmetry
breaking, spontaneous or explicit, does not necessarily imply
remnant symmetry breaking.


\section{Conclusions}\label{conclusions}

    Results of our numerical simulations suggest an appealing
picture: The low-lying eigenvalues of the Faddeev--Popov operator in
Coulomb gauge tend towards zero as the lattice volume increases. The
density of the eigenvalues goes as a small power of $\lambda$, and
this, together with a similar behavior of the average Laplacian,
$F(\lambda)$, assures the infrared divergence of the energy of an
unscreened color charge. Also, due to the enhancement of near-zero
modes of the Faddeev--Popov operator, the Coulomb energy of a pair
of color charges rises linearly with their separation. Both facts
support the ideas of the Gribov-horizon confinement scenario.

    The constant density of low-lying eigenvalues can be
attributed to the vortex component of gauge-field configurations. A
thermalized configuration in a pure gauge theory factors into a
confining piece (the vortex-only part), and a piece which closely
resembles the lattice of a gauge--Higgs theory in the Higgs phase
(the vortex-removed configuration). This establishes a firm
connection between the center-vortex picture and the Gribov-horizon
scenario. This connection is exemplified also by the fact that
vortex removal removes the color Coulomb string tension of the color
Coulomb potential. It is also consistent with recent investigations
of Gattnar et al.\ \cite{Gattnar:2004bf} in Landau gauge.

    In this talk, I covered results of our numerical investigations.
Related analytical developments were omitted: (i) thin center
vortices lie on the Gribov horizon; (ii) the Gribov horizon is a
convex manifold in lattice configuration space and thin center
vortices are conical singularities on that manifold; (iii) the
Coulomb gauge is an attractive fixed point of a more general gauge
condition, interpolating between the Coulomb and Landau gauges.
Interested readers are invited to find them in our recent
publications~\cite{Greensite:2004ke,Greensite:2004ur}.


\begin{theacknowledgments}
\begin{compacttext}
{Our research is supported in part by the U.S. Department of Energy
under Grant No.\ DE-FG03-92ER40711 (J.G.), the Slovak Grant Agency
for Science, Grant No. 2/3106/2003 (\v{S}.O.), and the National
Science Foundation, Grant No. PHY-0099393 (D.Z.). \v{S}.O.\ is
grateful to the organizers of the conference for invitation to
present this talk and for creating a stimulating, yet relaxed
atmosphere.\phantom{\v{S}}}
\end{compacttext}
\end{theacknowledgments}



\begin{thebibliography}{23}
\expandafter\ifx\csname
natexlab\endcsname\relax\def\natexlab#1{#1}\fi
\providecommand{\enquote}[1]{``#1''} \expandafter\ifx\csname
url\endcsname\relax
  \def\url#1{\texttt{#1}}\fi
\expandafter\ifx\csname urlprefix\endcsname\relax\def\urlprefix{URL
}\fi

\bibitem[Saxe(1816-1887)]{Saxe:1887}
Saxe, J.~G. (1816--1887), {\it{The Blind Men and the Elephant}}.

\bibitem[Greensite(2003)]{Greensite:2003bk}
Greensite, J., \emph{Prog.\ Part.\ Nucl.\ Phys.}, \textbf{51}, 1
(2003), hep-lat/0301023.

\bibitem[Gribov(1978)]{Gribov:1977wm}
Gribov, V.~N., \emph{Nucl.\ Phys.}, \textbf{B139}, 1 (1978).

\bibitem[Zwanziger(1998)]{Zwanziger:1998ez}
Zwanziger, D., \emph{Nucl.\ Phys.}, \textbf{B518}, 237 (1998).

\bibitem[Greensite and {Olejn{\'\i}k}(2003)]{Greensite:2003xf}
Greensite, J., and {Olejn{\'\i}k}, {\v{S}}., \emph{Phys.\ Rev.},
\textbf{D67}, 094503 (2003), hep-lat/0302018.

\bibitem[Greensite et~al.(2004{\natexlab{a}})]{Greensite:2004ke}
Greensite, J., {Olejn{\'\i}k}, {\v{S}}., and Zwanziger, D.,
\emph{Phys.\ Rev.}, \textbf{D69}, 074506 (2004{\natexlab{a}}),
hep-lat/0401003.

\bibitem[Greensite et~al.(2004{\natexlab{b}})]{Greensite:2004ur}
Greensite, J., {Olejn{\'\i}k}, {\v{S}}., and Zwanziger, D.
  (2004{\natexlab{b}}), hep-lat/0407032.

\bibitem[Christ and Lee(1980)]{Christ:1980ku}
Christ, N.~H., and Lee, T.~D., \emph{Phys.\ Rev.}, \textbf{D22}, 939
(1980).

\bibitem[Dell'Antonio and Zwanziger(1991)]{Dell'Antonio:1991xt}
Dell'Antonio, G., and Zwanziger, D., \emph{Commun.\ Math.\ Phys.},
\textbf{138}, 291 (1991).

\bibitem[Del~Debbio et~al.(1997)]{DelDebbio:1996mh}
Del~Debbio, L., Faber, M., Greensite, J., and {Olejn{\'\i}k},
{\v{S}}., \emph{Phys.\ Rev.}, \textbf{D55}, 2298 (1997),
hep-lat/9610005.

\bibitem[Del~Debbio et~al.(1998)]{DelDebbio:1998uu}
Del~Debbio, L., Faber, M., Giedt, J., Greensite, J., and
{Olejn{\'\i}k}, {\v{S}}., \emph{Phys.\ Rev.}, \textbf{D58}, 094501
(1998), hep-lat/9801027.

\bibitem[de~Forcrand and D'Elia(1999)]{deForcrand:1999ms}
de~Forcrand, P., and D'Elia, M., \emph{Phys.\ Rev.\ Lett.},
\textbf{82}, 4582 (1999), hep-lat/9901020.

\bibitem[Lang et~al.(1981)]{Lang:1981qg}
Lang, C.~B., Rebbi, C., and Virasoro, M., \emph{Phys.\ Lett.},
\textbf{B104}, 294 (1981).

\bibitem[Osterwalder and Seiler(1978)]{Osterwalder:1977pc}
Osterwalder, K., and Seiler, E., \emph{Ann.\ Phys.}, \textbf{110},
440 (1978).

\bibitem[Fradkin and Shenker(1979)]{Fradkin:1978dv}
Fradkin, E.~H., and Shenker, S.~H., \emph{Phys.\ Rev.},
\textbf{D19}, 3682 (1979).

\bibitem{Chernodub:1998wh}
Chernodub, M.~N., Gubarev, F.~V., Ilgenfritz, E.~M., and Schiller,
A., \emph{Phys.\ Lett.}, \textbf{B434}, 83 (1998), hep-lat/9805016.

\bibitem[Langfeld(2001)]{Langfeld:2001he}
Langfeld, K. (2001), hep-lat/0109033.

\bibitem[Bertle et~al.(2004)]{Bertle:2003pj}
Bertle, R., Faber, M., Greensite, J., and {Olejn{\'\i}k}, {\v{S}}.,
\emph{Phys.\ Rev.}, \textbf{D69}, 014007 (2004), hep-lat/0310057.

\bibitem[Zwanziger(2003)]{Zwanziger:2002sh}
Zwanziger, D., \emph{Phys.\ Rev.\ Lett.}, \textbf{90}, 102001
(2003), hep-lat/0209105.

\bibitem[Nakamura(2004)]{Nakamura:2004an}
Nakamura, A. (2004), \lowercase{t}alk at this conference.

\bibitem[Cucchieri and Zwanziger(2003)]{Cucchieri:2002su}
Cucchieri, A., and Zwanziger, D., \emph{Nucl.\ Phys.\ Proc.\
Suppl.}, \textbf{119}, 727 (2003), hep-lat/0209068.

\bibitem[Langfeld and Moyaerts(2004)]{Langfeld:2004qs}
Langfeld, K., and Moyaerts, L., \emph{Phys.\ Rev.}, \textbf{D70},
074507 (2004), hep-lat/0406024.

\bibitem[Greensite and Thorn(2002)]{Greensite:2001nx}
Greensite, J., and Thorn, C.~B., \emph{JHEP}, \textbf{02}, 014
(2002).

\bibitem[Gattnar et~al.(2004)]{Gattnar:2004bf}
Gattnar, J., Langfeld, K., and Reinhardt, H., \emph{Phys.\ Rev.\
Lett.}, \textbf{93}, 061601 (2004), hep-lat/0403011.

\end{thebibliography}

\end{document}